\documentclass[aps,prl,twocolumn,superscriptaddress]{revtex4-1}

\usepackage[latin9]{inputenc}
\usepackage{graphicx,color,graphics}
\usepackage{dcolumn}
\usepackage{amsmath}
\usepackage{amssymb}
\usepackage{amsfonts}
\usepackage{hyperref}
\usepackage{url}
\usepackage{bm}
\usepackage{lipsum}
\usepackage{xcolor}
\usepackage{ulem}
\makeatletter
\newcommand{\ket}[1]{{{|}{#1}\rangle}}

\newcommand{\plocal}{p_{\mathrm{local}}}
\newcommand{\pml}{p^{\mathrm{min}}_{\mathrm{local}}} 
\newcommand{\tsup}[1]{\textsuperscript{#1}}

\begin{document}

\title{Chained Bell Inequality Experiment with High-Efficiency Measurements} 

\author{T. R. Tan}
\email[Electronic address: ]{tingrei.tan@nist.gov}
\email{tingrei86@gmail.com}
\affiliation{National Institute of Standards and Technology, 325 Broadway, Boulder, Colorado 80305, USA}
\affiliation{Department of Physics, University of Colorado, Boulder, Colorado 80309, USA}
\author{Y. Wan}
\affiliation{National Institute of Standards and Technology, 325 Broadway, Boulder, Colorado 80305, USA}
\author{S. Erickson}
\affiliation{National Institute of Standards and Technology, 325 Broadway, Boulder, Colorado 80305, USA}
\affiliation{Department of Physics, University of Colorado, Boulder, Colorado 80309, USA}
\author{P. Bierhorst}
\author{D. Kienzler}
\author{S. Glancy}
\affiliation{National Institute of Standards and Technology, 325 Broadway, Boulder, Colorado 80305, USA}
\author{E. Knill}
\affiliation{National Institute of Standards and Technology, 325 Broadway, Boulder, Colorado 80305, USA}
\affiliation{Center for Theory of Quantum Matter, University of Colorado, Boulder, Colorado 80309, USA}
\author{D. Leibfried}
\author{D. J. Wineland}
\affiliation{National Institute of Standards and Technology, 325 Broadway, Boulder, Colorado 80305, USA}

\begin{abstract}
We report correlation measurements on two $^9$Be$^+$ ions that violate a chained Bell inequality obeyed by any local-realistic theory. The correlations can be modeled as derived from a mixture of a local-realistic probabilistic distribution and a distribution that violates the inequality. A statistical framework is formulated to quantify the local-realistic fraction allowable in the observed distribution without the fair-sampling or independent-and-identical-distributions assumptions. We exclude models of our experiment whose local-realistic fraction is above 0.327 at the 95 \% confidence level. This bound is significantly lower than 0.586, the minimum fraction derived from a perfect Clauser-Horne-Shimony-Holt inequality experiment. Furthermore, our data provides a device-independent certification of the deterministically created Bell states.
\end{abstract}

\maketitle

Recently several groups have reported loophole-free tests of local realism with Bell's theorem \cite{Bell1964}, rejecting with high confidence theories of local realism \cite{Hensen2015,Shalm2015,Giustina2015}. While these experiments falsify the idea that nature obeys local realism, they are limited in the extent to which their data differs from local realism. Chained Bell inequality (CBI) \cite{Pearle1970} experiments can show greater departures from local realism in the following sense:  Elitzur, Popescu, and Rohrlich \cite{Elitzur1992} described a model of the distribution of outcomes measured from a quantum state as a mixture of a local-realistic distribution, which obeys Bell's inequalities, and another distribution that does not.  Following their convention, we call these distributions ``local'' and ``non-local.'' According to Ref. \cite{Elitzur1992}, a probability distribution $P$ for the outcomes of an experiment can be written as 
\begin{equation}
P = \plocal P^{L} + (1-\plocal) P^{NL},
\label{LocalAndNonLocal}
\end{equation}
where $P^{L}$ represents a local joint probability distribution (a ``local part'') and $P^{NL}$ represents a non-local distribution, with $\plocal$ as the weight of the local component bound by $0\leq \plocal\leq1$. For an ideal Clauser-Horne-Shimony-Holt (CHSH) Bell inequality experiment where two physical systems (usually particles) are jointly measured with four different measurement settings \cite{CHSH1969}, the lowest attainable upper bound on the local content $\plocal$ in any quantum distribution is $\sim 0.586$ \cite{Barrett2006, Christensen2015}. In principle, this bound can be lowered to zero by using a chained Bell inequality experiment. 

\begin{figure}
\includegraphics[width=1\linewidth]{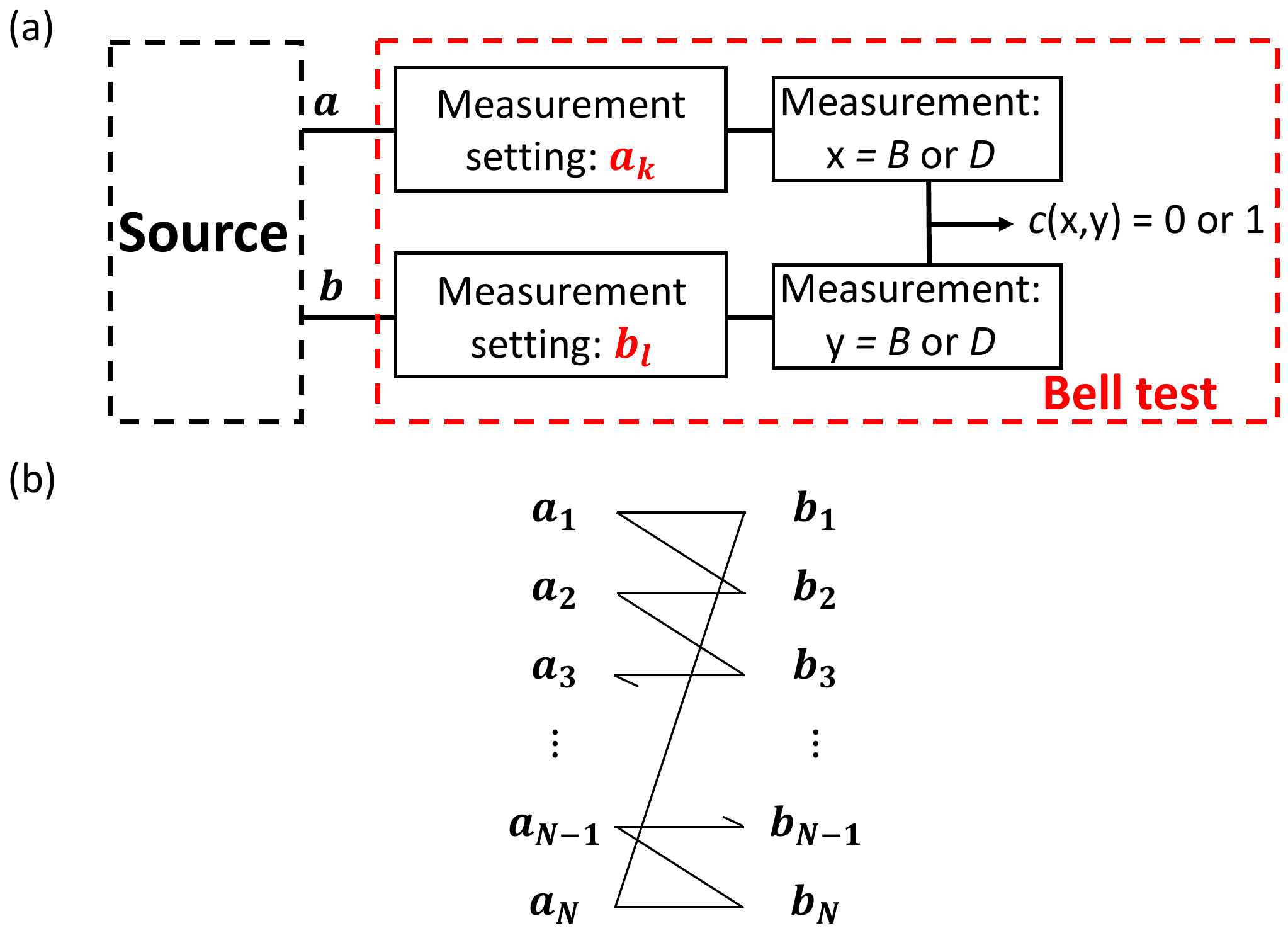}
\caption{(a) Illustration of a Bell inequality experiment. A source emits two systems $a$ and $b$, here two $^9$Be$^+$ ions. After choosing measurement settings $a_k$ and $b_l$, the experiment implements Hilbert-space rotation operations corresponding to these settings (which are controlled with classical variables) on the ions respectively. Then, a standard fluorescence based measurement in a fixed basis is applied to each ion. This is equivalent to choosing the measurement basis for the state that is present before the measurement settings are applied. Each system's measurement outcome is labeled $B$ for ``bright'' or $D$ for ``dark'', corresponding to the observation of fluorescence or not. From the joint measurement we record ``c = 1'' if the outcomes are the same and ``c = 0'' if they are not. (b) ``Chaining'' of the measurement settings for the $N$th CBI experiment. The measurement settings can be visualized as a chain where $a_kb_k$ and $a_{k+1}b_{k+1}$ are linked by $a_kb_{k+1}$, and the chain is closed by the settings $a_Nb_1$. The CHSH Bell inequality experiment corresponds to the special case of $N=2$. }
\label{BellTest}
\end{figure}

As indicated in Fig. \ref{BellTest}.(a), CBI experiments are a generalization of a CHSH-type experiment. During each trial, a source that may be treated as a ``black box'' emits two systems labeled $a$ and $b$, respectively. The experimentalist records the measurement outcomes after choosing a pair of measurements to perform separately on $a$ and $b$.  We use the symbols $a_{k}$, $b_{l}$ to denote the respective measurement settings and $a_{k}b_{l}$ for the pair. The latter is usually simply referred to as ``the settings'' or ``the setting pairs''. There is a hierarchy in which the $N$th CBI experiment involves $2N$ different settings. The $N=2$ CBI experiment is equivalent to the CHSH Bell inequality experiment. The settings for general $N$ are chosen from the set
\begin{equation}
Z=\{a_1b_1, a_1b_2, a_2b_2, a_2b_3, \ldots, a_{N-1}b_{N}, a_Nb_N, a_Nb_1\}. 
\label{MeasurementSet}
\end{equation}
Each local measurement has a binary outcome of $B$ for ``bright'' or $D$ for ``dark'' (Fig. \ref{BellTest}). The outcome of the trial is recorded as $c(x,y)= 1$ if $x=y $ or 0 if $x \neq y$, where $x$ is the outcome from system $a$ and $y$ is the outcome from system $b$. The probability to obtain $c(x,y)=1$ may depend on the choices $a_k$ and $b_l$, so we define that probability to be the correlation $\mathcal{C}(a_k,b_l)=P(BB|a_kb_l)+P(DD|a_kb_l)$, where $P(xy|a_kb_l)$ is the probability that system $a$ yields measurement outcome $x$ and system $b$ yields measurement outcome $y$ when the setting pair is $a_kb_l$. We define the chained Bell parameter to be
\begin{eqnarray}
I_N &=& \mathcal{C}(a_1,b_1)+\mathcal{C}(a_1,b_2)+\mathcal{C}(a_2,b_2)+...\nonumber\\
&&..+\mathcal{C}(a_N,b_N)+\left(1-\mathcal{C}(a_N,b_1)\right).\label{Iequation}
\end{eqnarray}
If the experiment is governed by a local hidden variable model, then the CBI $I_N \geq 1$ must be satisfied \cite{Pearle1970}. Note that $I_N$ can be estimated using only the record of the settings $a_kb_l$ and outcomes $c(x,y)$, without knowledge of the mechanism of the source. It was shown in Ref. \cite{Barrett2006} that the chained Bell parameter $I_N$ is always an upper bound on $\plocal$. In fact, $I_{N}$ is a least upper bound for $\plocal$ under the assumption that the distributions are non-signaling,  in the sense that each party's measurement outcomes do not depend on the other party's setting choice \cite{Bierhorst2016}. In the limit of $N\rightarrow\infty$ and with perfect experimental conditions, CBI experiments could be used to show that $\plocal\rightarrow 0$, demonstrating complete departure from local realism.

Similar to a CHSH-type experiment, a CBI experiment may be subject to ``loopholes'' \cite{Brunner2014, Larsson2014} that, in principle, allow local systems to show violation of the inequality. These loopholes arise when one must rely on various supporting assumptions that are made in the design and execution of the experiments, but which cannot be absolutely verified. For example, if the setting choice for $a$ can be communicated to $b$ (or vice-versa), the ``locality loophole'' is opened. Ensuring space-like separation between the choices and remote measurement events closes this loophole \cite{Bell1985}. The ``detection loophole'' \cite{Pearle1970, Clauser1974} is opened by making the fair-sampling assumption, which says that a subset of the data can be used to represent the entire data set.  This assumption is often used when some trials fail to produce outcomes due to inefficient detectors. High efficiency detectors are required to close the detection loophole and observe violation of the inequality \cite{Cabello2009}. The minimum detection efficiency required to close the detection loophole for the $N$th CBI experiment is given by Ref. \cite{Cabello2009} as
\begin{equation}
\eta_{min}(N)=\frac{2}{\frac{N}{N-1}\mathrm{cos}(\frac{\pi}{2N})+1},
\end{equation}
assuming that the measurement efficiencies on $a$ and $b$ are equal and that a maximally entangled state is measured. This emphasizes the importance of high detection efficiencny in large $N$ CBI experiments. If the analysis of the data assumes that the outcomes of the trials are independent and identically distributed (i.i.d.), the ``memory loophole'' is opened \cite{Barrett2002}. For example, one way to determine $I_N$ is by running each of the CBI setting pairs $a_kb_l$ for a total number of $M_{k,l}$ trials respectively and calculating
\begin{equation}
\overline{\mathcal{C}}(a_l,b_k)=\frac{\sum_{i=1}^{M_{k,l}} c(x_i,y_i)}{M_{k,l}},
\label{Cbar}
\end{equation}
(where $i$ indexes the trials) to estimate each $\mathcal{C}(a_l,b_k)$ term in Eq. (\ref{Iequation}). This analysis requires the i.i.d.\ assumption for standard error estimates to be valid. The memory loophole can be closed by applying appropriate analysis techniques to an experiment that uses randomized settings for each trial \cite{Gill2003}.

Previous experiments on the CBIs employed entangled photons pairs \cite{Pomarico2011,Aolita2012,Stuart2012,Christensen2015}. The lowest yet reported upper bound on $\plocal$ is approximately 0.126 for $N = 18$ \cite{Christensen2015}. However, to our knowledge all previous CBI experiments with $N \geq 3$ are open to the locality, detection, and memory loopholes.

Here, with a pair of atomic ions, we experimentally put an upper bound on $\plocal$ by measuring $I_N$ with near 100 \% detection efficiency. The measurement outcomes of every trial in each experiment are recorded and used to determine $I_N$, so the detection loophole is closed, as first incorporated in a CHSH Bell inequality experiment \cite{Rowe2001}. Furthermore, we address the memory loophole in a $N=6$ CBI experiment by employing uniformly random settings and developing a statistical analysis technique that does not require the assumption that trials are i.i.d. However, with each ion's measurement inside the lightcone of the event where the other ion's setting choice is made, we do not close the locality loophole.

\begin{figure}
\includegraphics[width=1\linewidth]{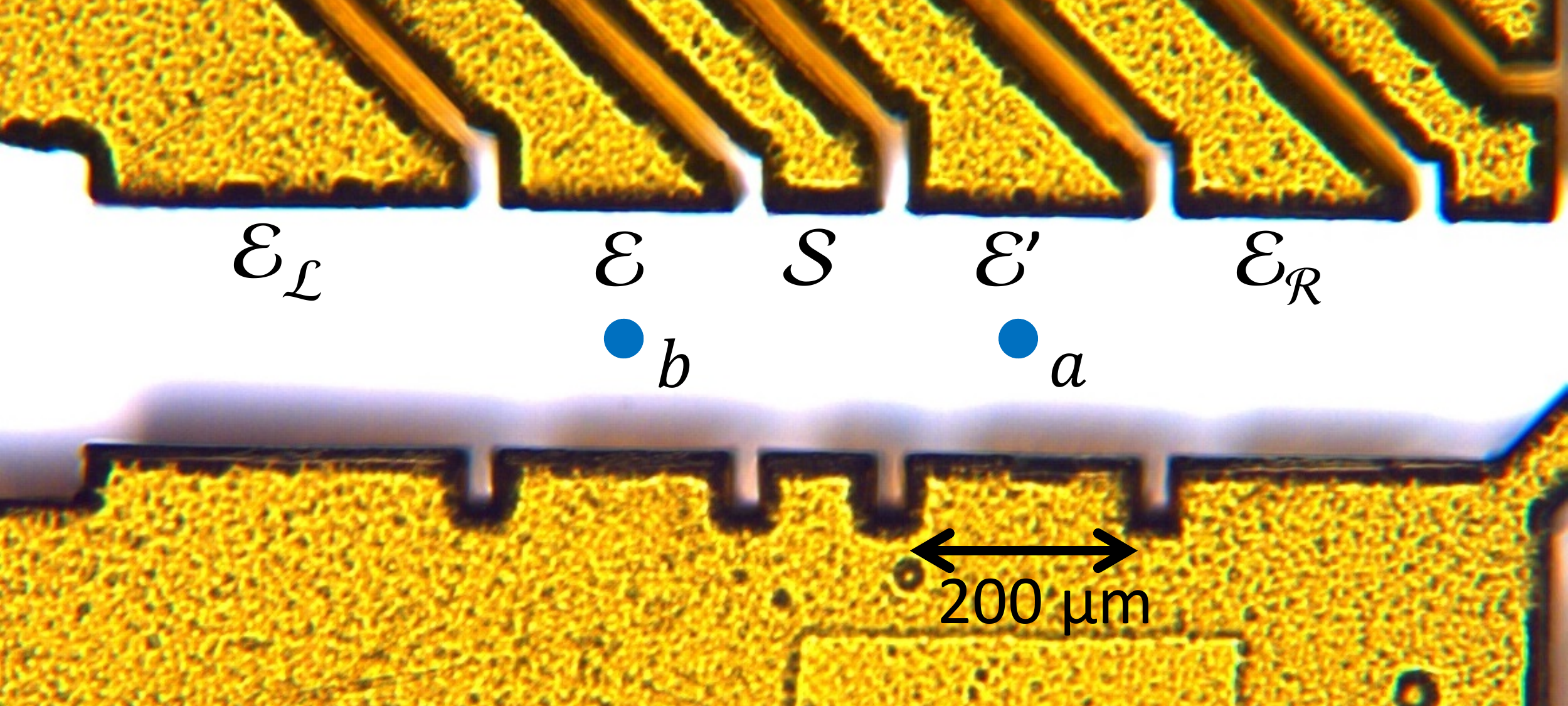}
\caption{Layout of the relevant segmented trap electrodes. Each CBI experiment begins with one ion located in zone $\mathcal{E}$ and the other in zone $\mathcal{E'}$. The blue dots, which indicate the ions, are overlaid on a photograph showing the trap electrodes (gold). By transporting the ions in and out of zone $\mathcal{S}$, we individually implement settings and measure each ion sequentially (details in supplemental material). The ions are separated by at least $\sim 340$ $\mu$m when settings $a_kb_l$ are applied, a distance much larger than the laser beams size of $\sim$ 25 $\mu$m. }
\label{TrapElectrode}
\end{figure}

Two beryllium ions ($^9$Be$^+$) are confined and aligned along the axis of a linear Paul trap by applying a combination of radio frequency (RF) and static potentials \cite{Gaebler2016} (see Fig. \ref{TrapElectrode}). This trap features segmented control electrodes allowing ions to be confined in different wells by applying controlled potentials \cite{Blakestad2011}. The ions can be confined together in a single harmonic well, or separately confined in different locations along the trap axis. Time varying potentials are applied to the control electrodes to deterministically separate ions and transport them between different locations \cite{Blakestad2011,Bowler2012}. 

The two states of the ions are encoded in the two electronic ground-state hyperfine levels $\ket{F=2, m_F = 0}  = \ket{\downarrow}$ and $\ket{F=1,m_F=1} = \ket{\uparrow}$, where $F$ and $m_F$ are the total angular momentum and its projection along the quantization axis provided by an external magnetic field of $\simeq 0.0119$ T. The frequency splitting of the two states is approximately 1.2 GHz and is first-order insensitive to magnetic field fluctuations \cite{Langer2005}. With coherent operations based on stimulated-Raman transitions (with laser wavelengths near 313 nm), we can deterministically create the entangled states 
\begin{eqnarray}
\ket{\Phi_{+/-}}=\frac{1}{\sqrt{2}}\left(\ket{\uparrow\uparrow}\pm\ket{\downarrow\downarrow}\right),
\label{BellStates}
\end{eqnarray}
with high-fidelity, where the notation $\ket{\uparrow\uparrow}$ denotes the two ion state $\ket{\uparrow}_a\ket{\uparrow}_b$ \cite{Gaebler2016} (see supplemental material for details on the Bell state generation). States $\ket{\Phi_{+/-}}$ are created with the ions are located in zone $\mathcal{S}$ (Fig. \ref{TrapElectrode}). This is followed by separating the ions and placing them in two separate potential wells, one located in zone $\mathcal{E}$ and one located in $\mathcal{E'}$, separated by $\sim$ 340 $\mu$m. These processes represent the source in Fig. \ref{BellTest} and prepare the  two ions $a$ and $b$ for the measurement of $I_N$ described below. 

To implement the different settings, we illuminate the ions with stimulated-Raman-transition-inducing laser beams controlled with classical parameters. Ideally, they can be described as the following rotations
\begin{eqnarray}
\ket{\uparrow}_{r}&\rightarrow&\frac{1}{\sqrt{2}}\left(\ket{\uparrow}_{r}-\mathrm{e}^{-i r_k}\ket{\downarrow}_{r}\right),\nonumber\\
\ket{\downarrow}_{r}&\rightarrow&\frac{1}{\sqrt{2}}\left(\ket{\downarrow}_{r}-\mathrm{e}^{i r_k}\ket{\uparrow}_{r}\right), 
\label{SingleQubitRotation}
\end{eqnarray}
where $r = a$ or $b$ to represent each of the ions, and the angles $r_k = a_k$ or $b_k$ are
\begin{eqnarray}
a_k &=& \frac{(2k-1)\pi}{2N},\\
b_l &=& -\frac{(l-1)\pi}{N},
\end{eqnarray}
which are chosen from Eq. (\ref{MeasurementSet}). These angles minimize the expected value of $I_{N}$ if the produced entangled state is ideal \cite{Braunstein1990}. These rotation operations are implemented by setting the amplitude and phase of the Raman laser beams with an acousto-optic modulator (AOM). The radio-frequency electric field driving the AOM is produced by a field-programmable gate array (FPGA)-controlled direct digital synthesizer. The classical variable is the phase of the oscillating field that implements a particular setting $a_k$. Analogous operations are applied to ion $b$ with setting $b_l$. The laser beams implementing these rotations have a beam waist of $\simeq 25$ $\mu$m and are focused at zone $\mathcal{S}$. They are applied sequentially to one of the ions in zone $\mathcal{S}$ while the other ion is located in a different well; each ion is transported in and out of zone $\mathcal{S}$ to interact with the laser beams (see supplemental material).

After the settings rotations are applied, the state of each ion, $\ket{\uparrow}$ or $\ket{\downarrow}$ is measured sequentially in zone $\mathcal{S}$ with a state-dependent fluorescence technique \cite{Blatt1988}. When the detection laser beam is applied, we detect on average 30 photon counts on a photomultiplier tube if the ion is in the $\ket{\uparrow}$ state and about 2 counts if the ion is in the $\ket{\downarrow}$ state. Our photon collection apparatus images ions in zone $\mathcal{S}$ with a field of view of approximately 50 $\mu$m. We label a measurement outcome ``dark'' ($D$) if 6 or fewer photons are observed and ``bright'' ($B$) if more than 6 are observed.  Thus we obtain the 4 possible joint-measurement fluorescence outcomes $BB$, $BD$, $DB$, $DD$, for each trial. These outcomes correspond to the states $\ket{\uparrow\uparrow}$, $\ket{\uparrow\downarrow}$, $\ket{\downarrow\uparrow}$, and $\ket{\downarrow\downarrow}$.  Among previous CHSH-type experiments with trapped ions \cite{Rowe2001,Matsukevich2008,Pironio2010,Ballance2015}, only two were performed with ions manipulated and measured in individual wells \cite{Matsukevich2008, Pironio2010}. In those experiments the ions were confined in two traps separated by about $\sim 1$ m. 

When the state $\ket{\Phi_{+}}$ is prepared, we compute an estimate $\widehat{I}_N$ of $I_N$ as shown in Eq. (\ref{Iequation}) with Eq. (\ref{Cbar}) used to estimate the $\overline{\mathcal{C}}(a_k,b_l)$ terms. For the state $\ket{\Phi_{-}}$ we instead use anticorrelations and compute
\begin{eqnarray}
\widehat{I}^\mathcal{A}_N &=& \overline{\mathcal{A}}(a_1,b_1)+\overline{\mathcal{A}}(a_1,b_2)+\overline{\mathcal{A}}(a_2,b_2)+...\nonumber\\
&&..+\overline{\mathcal{A}}(a_N,b_N)+\left(1-\overline{\mathcal{A}}(a_N,b_1)\right),\label{Iequation2}
\end{eqnarray}
where $\overline{\mathcal{A}}(a_k,b_l)= 1-\overline{\mathcal{C}}(a_k,b_l)$. The measured $\widehat{I}^\mathcal{A}_N$ is equivalent to $\widehat{I}_N$ for the purpose of quantifying $\plocal$. 

\begin{figure}
\includegraphics[width=1\linewidth]{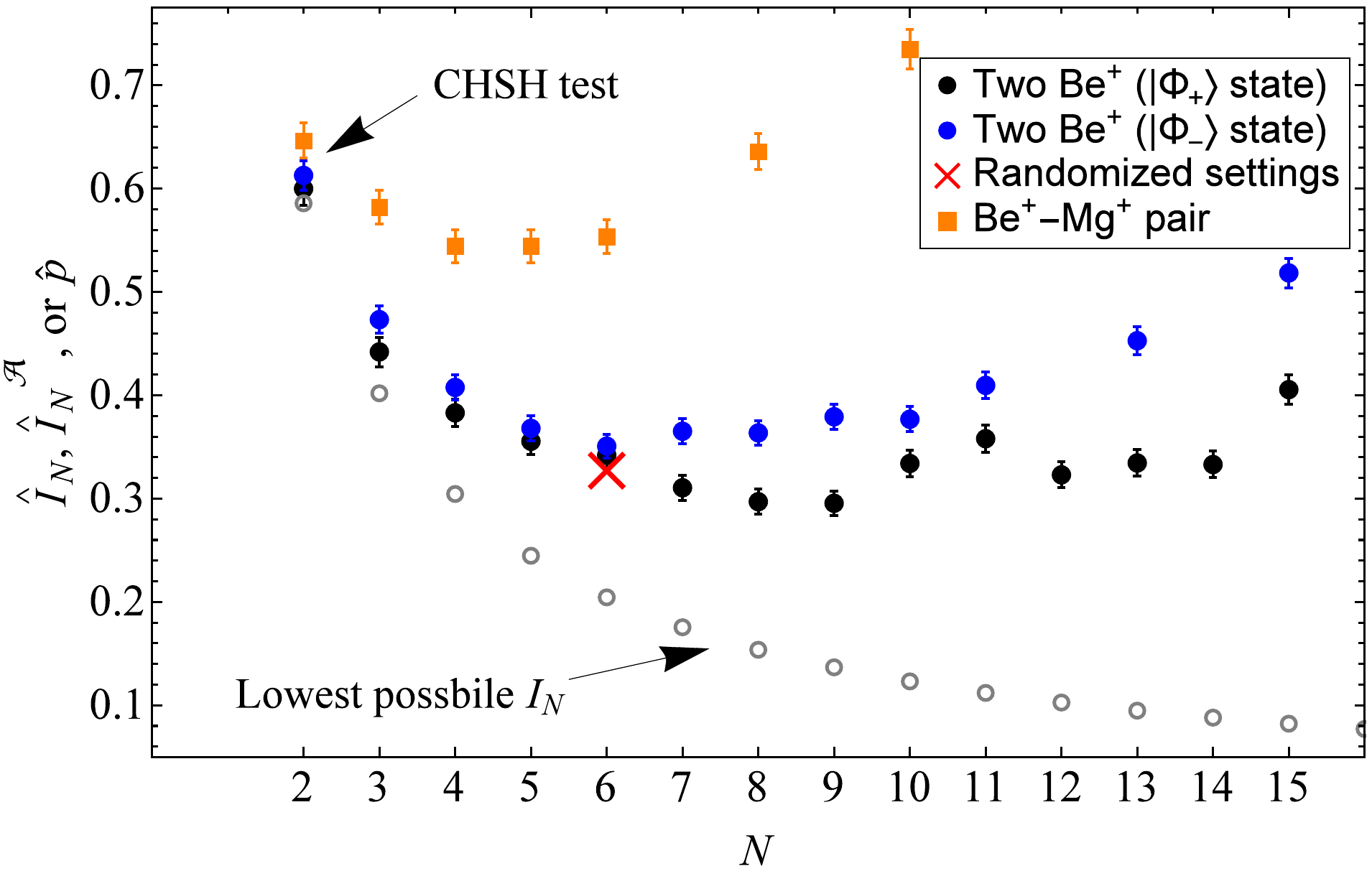}
\caption{Experimentally measured values $\widehat{I}_N$ and $\widehat{I}^{\mathcal{A}}_N$ as a function of $N$. Data represented by black and blue dots are obtained with two $^9$Be$^+$ ions, with black (blue) dots corresponds to tests on $\ket{\Phi_{+}}$ ($\ket{\Phi_{-}}$). These two data sets were obtained approximately six months apart. The difference between them and the finer features within each data set are probably due to miscalibrations and our inability to reproduce exact experimental conditions. Orange squares are data from tests on $\ket{\Phi_{+}}$ prepared on a $^9$Be$^+$-$^{25}$Mg$^+$ pair. The red cross represents the (95 \% confidence level) upper bound $\widehat{p} = 0.327$, estimated using our statistical framework, which does not require the i.i.d.\ assumption. The gray circles indicate the lowest $I_N$ achievable with perfect CBI experiments using a maximally entangled state. As $N$ increases from 2, the experimentally measured values of $\widehat{I}_N$ for each different pair of ions reach a minimum and then trend upward. This is due to errors that accumulate during experiments with higher values of $N$. In general, $\widehat{I}_N$ becomes more sensitive to errors and noise as $N$ increases \cite{Cabello2009}. }
\label{ChainedBellTest}
\end{figure}

We performed the experiment for the CBI parameter $N$ ranging from 2 to 15. Two different data sets, collected $\sim 6$ months apart, were obtained. Figure \ref{ChainedBellTest} shows the experimentally obtained CBI parameter $\widehat{I}_N$ as a function of $N$. The data points in Fig. \ref{ChainedBellTest} were obtained with multiple sequential trials having the same settings, then iterated across different choices of settings. The error bars are calculated under the assumption that the settings and outcomes are i.i.d. The error bars indicate the propagated standard errors $\sqrt{\sum_{j} \epsilon_j^2}$, with $\epsilon_j = \sqrt{\chi_j(1-\chi_j)/(M_j-1)}$ where $M_j$ is the number of trials (here $M_j\sim 2,000$) and $\chi_j$ is the averaged correlated or anticorrelated outcome for the $j$th setting pair. The $\widehat{I}_{2}$ experiment took a total of $\sim 5$ minutes, the one for $\widehat{I}_{15}$ $\sim 20$ minutes. The lowest value of $\widehat{I}_N$ is obtained for the $N=9$ data run, which corresponds to $\widehat{I}_9=0.296(12)$.

To remove the i.i.d.\ assumption, we performed an $N=6$ experiment employing uniformly random settings. The settings were chosen with a pseudo-random generator during run time. For each randomly chosen setting pair, blocks of 100 trials with identical settings were carried out before changing to the next randomly selected setting pair. This procedure was repeated $1,398$ times. This is the number of blocks we obtained in a single day's experiment run and was deemed sufficient for our statistical analysis to be reasonably informative. While we could have run one trial per settings choice, this would have implied a low data collection efficiency since a single trial takes $\sim 10\;\mathrm{ms}$, but reprogramming the FPGA controlling the apparatus to change the settings takes $\sim 4\;\mathrm{s}$. This $N=6$ experiment took $\sim 7$ hours. Although 100 outcome pairs are available for each random settings choice, only a single trial from each block should be analyzed when not making the i.i.d.\ assumption. We chose ahead of time to use the center trial (the 50th trial of of each block) in our analysis. The choice of the 50th trial was arbitrary; any other choice would have also produced a valid analysis. To enable this choice, we assume that the 50th trial does not depend on the earlier trials in each block. Collisions between ions and background gases can cause the ions to overheat or be ejected from the potential well. To reduce the consequences of these effects, we checked the status of the ions with fluorescence measurements. If the measurements made prior to the beginning of each trial do not detect a problem with the ions, that trial was ``heralded'' and included in the analysis (see supplemental material). Because we use only information gained prior to the beginning of a trial to herald that trial, the detection loophole is not opened. When studying the 50th trial of each block, 1,361 trials were therefore analyzed.

A memory-robust statistical framework is formulated to infer a bound on the maximum local content in the observed correlation. For $\ket{\Phi_{+}}$, we draw inferences based on the statistic $T_i(x,y,a_k,b_l)$, defined as 
\begin{equation}
T_i(x,y,a_k,b_l)=\begin{cases}
0 & \text{if } x=y \text{ and } (a_k,b_l) \neq (a_N,b_1) \\
1 & \text{if } x \neq y \text{ and } (a_k,b_l) \neq (a_N,b_1) \\
1 & \text{if } x = y \text{ and } (a_k,b_l)=(a_N,b_1) \\
0 & \text{if } x \neq y \text{ and } (a_k,b_l)=(a_N,b_1) \\
\end{cases},
\label{eq:ti}
\end{equation}
where $x$ and $y$ are the measurement outcomes from the two ions when they are measured with settings $a_kb_l$ chosen from Eq. (\ref{MeasurementSet}) during trial $i$. As a trial-by-trial function of both settings and outcomes, $T_i$ is more suitable for memory-robust statistical analysis than a statistic (such as $\overline{\mathcal{C}}$) that is normalized by the number of times each measurement setting occurs \cite{Barrett2002}. The expectation value of $T_i$ is $1-I_N/(2N)$, so intuitively larger values of $\sum_{i=1}^{1,361}T_i$ should correspond to a lower local fraction $P^{L}$ in Eq. (\ref{LocalAndNonLocal}). In the presence of memory effects, it is possible for the proportion of local states to change over time. Hence we model each trial $i$ as having a probability $p^i_{\mathrm{local}}$ of generating a local state, and derive a one-sided $1-\alpha$ confidence interval $[0,\widehat{p}]$ for $\pml:=\min_ip^i_{\mathrm{local}}$, the minimum local content that can occur over the course of the experiment. The probability of seeing a $\sum_{i=1}^nT_i$ statistic as large or larger than that actually observed for a model with $\pml=q$ is less than or equal to the same probability for an i.i.d.\ model with $p^i_{\mathrm{local}}=q$ for all $i$. From this, the desired confidence interval can be obtained by inversion of hypothesis-test acceptance regions (\S 7.1.2 of~\cite{Shao2003}); see the supplemental material.  In particular, this implies that we can take $\widehat{p}$ to be the largest value $x$ for which a binomial random variable with $n=1,361$ trials and probability of success $(2N-x)/2N$ yields a value as great or greater than the observed value of $\sum_iT_i$ with a probability of at least $\alpha$.

We compute one-sided $\pml$ confidence intervals of [$0,0.327$], [$0,0.366$], and [$0,0.413$] for confidence levels of 0.95, 0.99, and 0.999, respectively, using the 50th trial in each block of the randomized $N=6$ experiment. Randomization of settings and the use of this statistical framework also remove any concern that experimental drifts might erroneously lead to a lower estimate of $\pml$. 

The values of $\widehat{I}_6$ and $\widehat{p}$ obtained when we analyzed the 1st through 100th trial in each block, representing the $I_6$ estimates and confidence intervals that would have been obtained with a different choice of representative trial, are shown in the supplemental material.  When all data from the heralded trials is analyzed together, $\widehat{I}_6$ is determined to be 0.315(5). An associated confidence interval for local content would be valid only with additional assumptions. This is because the settings in each block are not random after the first. If the trials in each block were i.i.d., then one could use all 100 trials to create a confidence interval for local content. However, in our experiments the trials are not fully i.i.d.

Using the same apparatus, we also perform the CBI experiment on an entangled pair of $^9$Be$^+$ and $^{25}$Mg$^+$ ions. The computed $\widehat{I}_N$ values are shown as orange squares in Fig. \ref{ChainedBellTest}. The generation of the mixed-species entangled state is described in Ref. \cite{Tan2015}. In this experiment, the ions remain confined in a single zone throughout the entire sequence. This is because the laser wavelengths and the microwave frequencies used for manipulating the ions are sufficiently different that addressing one species with control fields negligibly affects the other species. The rotations implementing the settings choices are applied with microwave fields tuned to each ion. The frequencies are $\simeq$ 1.2 GHz for the $^9$Be$^+$ ion and $\simeq$ 1.8 GHz for the $^{25}$Mg$^+$ ion. The determinations of the final states of the two ions are made with detection lasers at $\sim 313\;\mathrm{nm}$ for the $^9$Be$^+$ and at $\sim 280\;\mathrm{nm}$ for the $^{25}$Mg$^+$ ion.

Our lowest measurement of $\widehat{I}_{2}$ corresponds to a CHSH inequality parameter (sum of correlations) of $B_{\mathrm{CHSH}}=2.80(2)$. Under local-realism $B_{\mathrm{CHSH}}=2\left(1-I_2\right)+2\leq 2$. A consequence of the near-maximal violations of the CHSH inequality ($B_{\mathrm{CHSH}}^{\mathrm{max}}=2\sqrt{2}\simeq 2.82$) provides a black box certification of the created entangled states \cite{Bardyn2009,Bancal2015,Kaniewski2016}. Such a characterization with minimal assumptions on our physical system and measurements is formalized by the self-testing framework \cite{Mayers2004,McKague2012}. Using the method of \cite{Kaniewski2016}, we infer a self-tested Bell-state fidelity lower bound (at the 95 \% confidence level) of $\sim 0.958$.

Our experiment is the first to report violation of CBIs for $N \geq 3$ while closing the detection loophole. Several previous experiments have reported violation of the CHSH inequality (the CBI with $N = 2$) while closing the detection loophole \cite{Rowe2001,Matsukevich2008,Pironio2010,Ballance2015,Tan2015,Ansmann2009,Hofmann2012,Pfaff2013,Vlastakis2015,Dehollain2016}, and \cite{Hensen2016} closed all loopholes (other experiments have closed the detection loophole, but they reported violation of other Bell inequalities, for which self-tested fidelity bounds are not available). Of these previous experiments, the largest reported CHSH parameters were 2.70(2) \cite{Tan2015} and 2.70(9) \cite{Dehollain2016}. These numbers yield self-tested Bell-state fidelity lower bounds (at the 95 \% confidence level according to Ref. \cite{Kaniewski2016}) of $\sim 0.888$ and $\sim 0.809$ respectively.

The apparatus used here is designed for the implementation of quantum information processing (QIP) with trapped-ions in a scalable system of trap zones in an array \cite{bible,Kielpinski2002}. The basic QIP elements incorporated for the realization of the CBI experiment described here include high-fidelity state preparation, manipulation, and measurement on individual qubits, transport between zones, qubits with long coherence times, and high-fidelity two-qubit gates. Therefore, CBI experimental results can also be regarded as a useful benchmark towards the goal of general purpose scalable QIP. 

In summary, our CBI results enabled us to reject models of our experiment in which the fraction of local distributions always exceeds 0.327, at the 95 \% confidence level. Furthermore, for the special case of the CHSH inequality, our self-tested fidelity appears to be the highest for a deterministically created Bell state.

\section{Acknowledgments}

This work was supported by the Office of the Director of National Intelligence (ODNI) Intelligence Advanced Research Projects Activity (IARPA), ONR and the NIST Quantum Information Program. We thank L. Shalm and J. Bergquist for helpful suggestions on the manuscript. We express gratitude to J. P. Gaebler, Y. Lin, R. J\"{o}rdens, R. Bowler, and A. C. Wilson for contributing to the experimental setup. S. Erickson is supported by the National Science Foundation Graduate Research Fellowship under Grant no. DGE 1650115. D. Kienzler acknowledges support from the Swiss National Science Foundation under grant no. 165208. This manuscript is a contribution of NIST and not subject to U.S. copyright.

\section{Supplemental Material}

\subsection{Generation of Bell states}

\begin{figure*}
\centering
\includegraphics[width=0.95\linewidth]{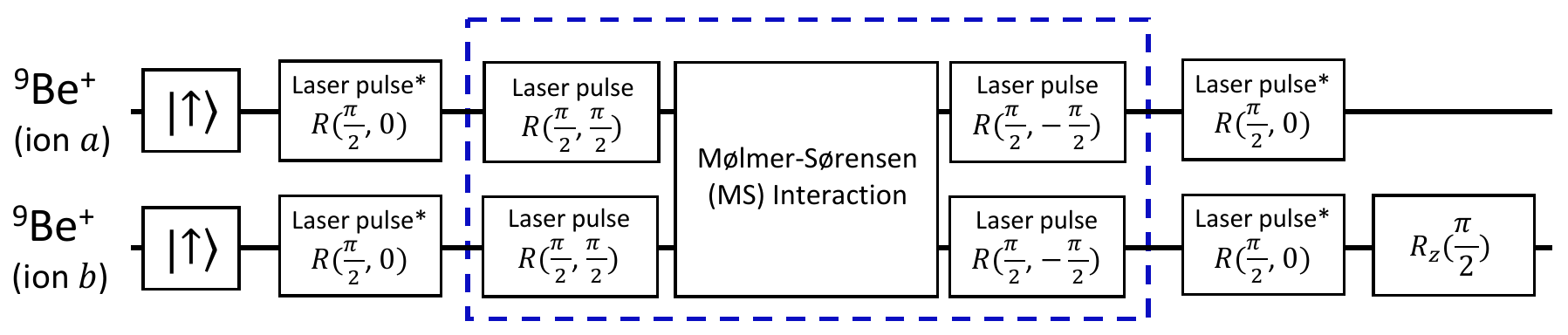}
\caption{Pulse sequence for the generation of the entangled state $\ket{\Phi_+}=\frac{1}{\sqrt{2}}\left(\ket{\uparrow\uparrow}+\ket{\downarrow\downarrow}\right)$. The notation $R(\theta,\phi)$ represents the rotation with angle $\theta$ about an axis in the $x$-$y$ plane of the Bloch sphere, and $\phi$ is the azimuthal angle of the rotation axis. Operation $R_z(\xi)$ is the rotation with an angle $\xi$ around the $z$ axis of the Bloch sphere. The angle $\theta$ is adjusted by varying the length of the laser pulse, and $\phi$ and $\xi$ are adjusted with the phases of the RF signal driving the AOMs that controls the laser beams. Laser pulses in the dashed box all use the same set of laser beams, which makes the sensitivity of the created state to slow phase drifts between the two Raman beam paths negligible \cite{Lee2005,Tan2015}. The two laser beams used to drive the stimulated-Raman transitions for the pulses outside of the dashed box (marked with asterisk) are copropagating which eliminates phase drifts due to path length differences in the beams \cite{Gaebler2016}. The $R_z$ rotation is implemented by shifting the direct digital synthesizer controlling the laser pulses that implement the measurement settings $b_l$.}
\label{BellStateGeneration}
\end{figure*}

To deterministically create a Bell state, we employ a pulse sequence (Fig. \ref{BellStateGeneration}) that consists of a two-qubit entangling gate and multiple global rotations induced by laser beams. The entangling gate is implemented with an effective M\o lmer-S\o rensen (MS) spin-spin interaction \cite{Sorensen1999}. The implementation and characterization of this logic gate is detailed in Ref. \cite{Gaebler2016}. 
The pulse sequence shown in the dashed box in Fig. \ref{BellStateGeneration} embeds the MS interaction in a Ramsey sequence, which reduces errors caused by slow phase drifts between Raman beams to negligible levels compared to other error sources \cite{Lee2005,Tan2015} and implements the two-qubit phase gate $\ket{\uparrow\uparrow}\rightarrow\ket{\uparrow\uparrow}$, $\ket{\uparrow\downarrow}\rightarrow i\ket{\uparrow\downarrow}$, $\ket{\downarrow\uparrow}\rightarrow i\ket{\downarrow\uparrow}$, and $\ket{\downarrow\downarrow}\rightarrow\ket{\downarrow\downarrow}$. With both ions initialized to the $\ket{\uparrow\uparrow}$ state, the overall pulse sequence in Fig. \ref{BellStateGeneration} ideally creates the entangled state $\ket{\Phi_+}=\frac{1}{\sqrt{2}}\left(\ket{\uparrow\uparrow}+\ket{\downarrow\downarrow}\right)$. From this state, we can effectively create the $\ket{\Phi_-}$ state by appropriately shifting the phase of the pulses that implement the measurement settings $b_l$. This is equivalent to a $\pi/2$-rotation around the $z$-axis of the Bloch sphere.

By measuring the population and the coherence of the created Bell state \cite{Sackett2000}, we determine the fidelity of the overall sequence in Fig. \ref{BellStateGeneration} to be approximately 0.99. The Bell state fidelity is lower than that reported in Ref. \cite{Gaebler2016} due to a higher error dominated by spontaneous emission of photons induced by the Raman laser beams \cite{Ozeri2007}. Here, different laser parameters, including Raman detuning, laser intensities, and polarizations compared to those used in Ref. \cite{Gaebler2016} are chosen to allow different operations (e.g. two-qubit gates, global rotations, and single-qubit rotations) to be implemented with the same laser beam lines. Furthermore, in Ref. \cite{Gaebler2016} the entangling gate was applied with ions confined in zone $\mathcal{E}$ (Fig. \ref{TrapElectrode}), but here entanglement is created in $\mathcal{S}$. We measure a factor of $\sim 2$ higher axial motional heating \cite{Turchette2000} rate in $\mathcal{S}$ than in $\mathcal{E}$. All of these factors reduce the fidelity of the entangling gate that creates the Bell states compared to Ref. \cite{Gaebler2016}.

\subsection{Transport of ions for separate rotations and measurements}

\begin{figure*}
\includegraphics[width=1\linewidth]{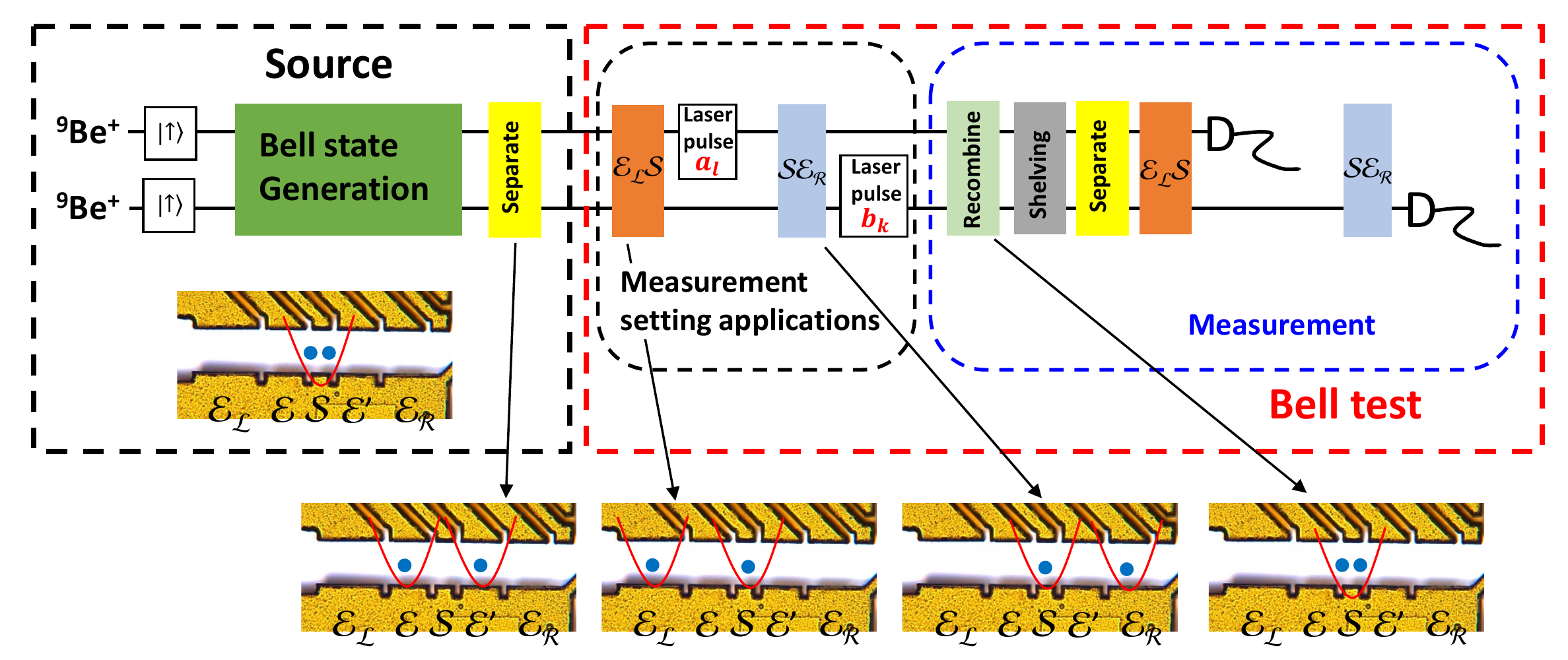}
\caption{Experimental sequence for one trial. The notation $\mathcal{E_LS}$ refers to transport to place the ion $b$ in zone $\mathcal{E_L}$ and ion $a$ in zone $\mathcal{S}$. Similarly for the operation $\mathcal{SE_R}$. The entangled state is generated as shown in Fig. \ref{BellStateGeneration} with the ions located in zone $\mathcal{S}$. Time-varying potentials are applied to control electrodes for the separation, shuttling and recombination of the ions \cite{Blakestad2011,Bowler2012}. }
\label{ExpSequence}
\end{figure*}

To apply the settings $a_k$ and $b_l$ to the ions individually, the ion in $\mathcal{E}$ is first transported to $\mathcal{E_L}$ while the ion in $\mathcal{E'}$ is simultaneously transported to $\mathcal{S}$ (see Fig. \ref{ExpSequence}). We then apply a laser beam to zone $\mathcal{S}$ to implement the measurement setting $a_k$ without disturbing ion $b$. Subsequently, time-varying voltages are applied to implement the simultaneous well transportation operations $\mathcal{S}\rightarrow\mathcal{E_R}$ and $\mathcal{E_L}\rightarrow\mathcal{S}$. With ion $b$ located in zone $\mathcal{S}$, a laser beam implements the measurement setting $b_l$. After applying these measurement settings, the ions are recombined into zone $\mathcal{S}$ for ``shelving'' pulses (see next section) to be applied on both ions simultaneously. Then, similar transport procedures separate and move the ions into zone $\mathcal{S}$ for individual fluorescence detection. 

\subsection{Ion state measurements}

\begin{figure}
\includegraphics[width=1\linewidth]{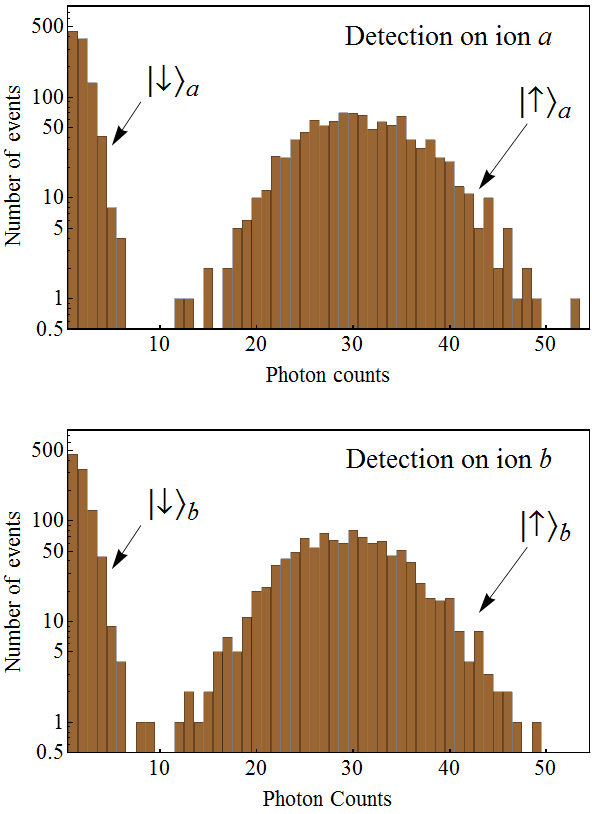}
\caption{Typical detection photon histograms obtained when individually exciting state-dependent fluorescence of ion $a$ and $b$. During detection of ion $a$ ($b$) at zone $\mathcal{S}$, ion $b$ ($a$) is located in zone $\mathcal{E_L}$ ($\mathcal{E_R}$) such that it does not interact with the detection laser beam. We choose a threshold of 6 for the differentiation of the ``bright'' and ``dark'' measurement outcomes.}
\label{IndividualDetection}
\end{figure}

We use state-dependent fluorescence to detect the ions' states.  This is accomplished by applying a $\sigma^+$-polarized laser beam to zone $\mathcal{S}$ on resonance with the $^2S_{1/2}\ket{2,2}$ to $^2P_{3/2}\ket{3,3}$ cycling transition for 330 $\mu$s \cite{Gaebler2016}. Before applying the detection laser, the population in the $\ket{1,1}$ state is transferred to the $\ket{2,2}$ state by a composite $\pi$ pulse driven by a microwave field. The population in the $\ket{2,0}$ state is first transferred to the $\ket{1,-1}$ state by a microwave $\pi$ pulse, then another $\pi$ pulse is applied to transfer any remaining population in the $\ket{2,0}$ state to the $\ket{1,0}$ state. These ``shelving'' microwave pulses are applied when the ions are recombined into zone $\mathcal{S}$ and are implemented to maximally distinguish the ``bright'' ($B$) and ``dark'' ($D$) states. After the application of these ``shelving'' pulses, the ions are separated into different potential wells to interact sequentially with the detection laser applied to zone $\mathcal{S}$. Figure \ref{IndividualDetection} shows the typical detection photon histograms of individual $^9$Be$^+$ ions. The detection error is experimentally estimated to be $\sim 3\times10^{-3}$ per ion. 
The procedure for measuring the $^{25}$Mg$^+$ ion is very similar to that of the $^{9}$Be$^+$ ion \cite{Tan2015}. Due to higher background photon counts when the CBI experiments were performed on the $^9$Be$^+$-$^{25}$Mg$^+$ pair, bright thresholds of 11 and 12 were used for $^9$Be$^+$ and $^{25}$Mg$^+$, respectively.

\subsection{Heralded collection of trials}

Due to the long duration of the randomized settings experiment, collisions of the ions with background gas can results in (i) high motional temperature, (ii) ejections of ions from the trapping potential, or (iii) formations of molecular ions \cite{bible}. High thermal excitations of the ions due to background gas collisions reduce the fidelity of the Bell state generated. Ejection of ions from the potential well and the formation of molecular ions render the ions useless. To reduce the effect of the background gas collisions on our estimates of $I_N$ and $\pml$, we determine the status of the ions by checking for fluorescence between trials by illuminating both ions with the detection laser beam. Collisions cause reductions of fluorescence photons during detection. A check that indicates both ions are likely to be present heralds the beginning of a trial that is included in our analysis. For the collection of the $q$th trial, we compute
\begin{equation}
H_{\mathrm{check}}^q=\sum_{f=q-g}^{q-1}H_f, 
\label{FluorescenceCheck}
\end{equation}
where $H_{\mathrm{check}}^q$ is the total photon counts for the determination of the ion's status in the $q$th trial, $H_f$ is the number of photon counts obtained in the $f$th fluorescence check, and $g$ is the number of fluorescence detection events to be included. The $q$th trial will be included for analysis if $H_{\mathrm{check}}^q > g H_{\mathrm{thres}}$. We use $H_{\mathrm{thres}}=20$ and $g=8$, these values are determined by analyzing detection histograms obtained with training data sets and with numerical simulations. 

The heralding method was designed after the acquisition of the data, but not modified based on the results of its application. For future experiments, it will be advantageous to increase the programming speed of the FPGA controlling the experimental setup and to reduce the collision rate of ions with background gases. For the non-randomized settings experiments, although a similar fluorescence checking procedure was also implemented, the data reported in the main text does not use heralding. When the heralding strategy is used, the analysis yields similar results (see next section).

\subsection{Example data and additional analysis}

\begin{figure}
\includegraphics[width=1\linewidth]{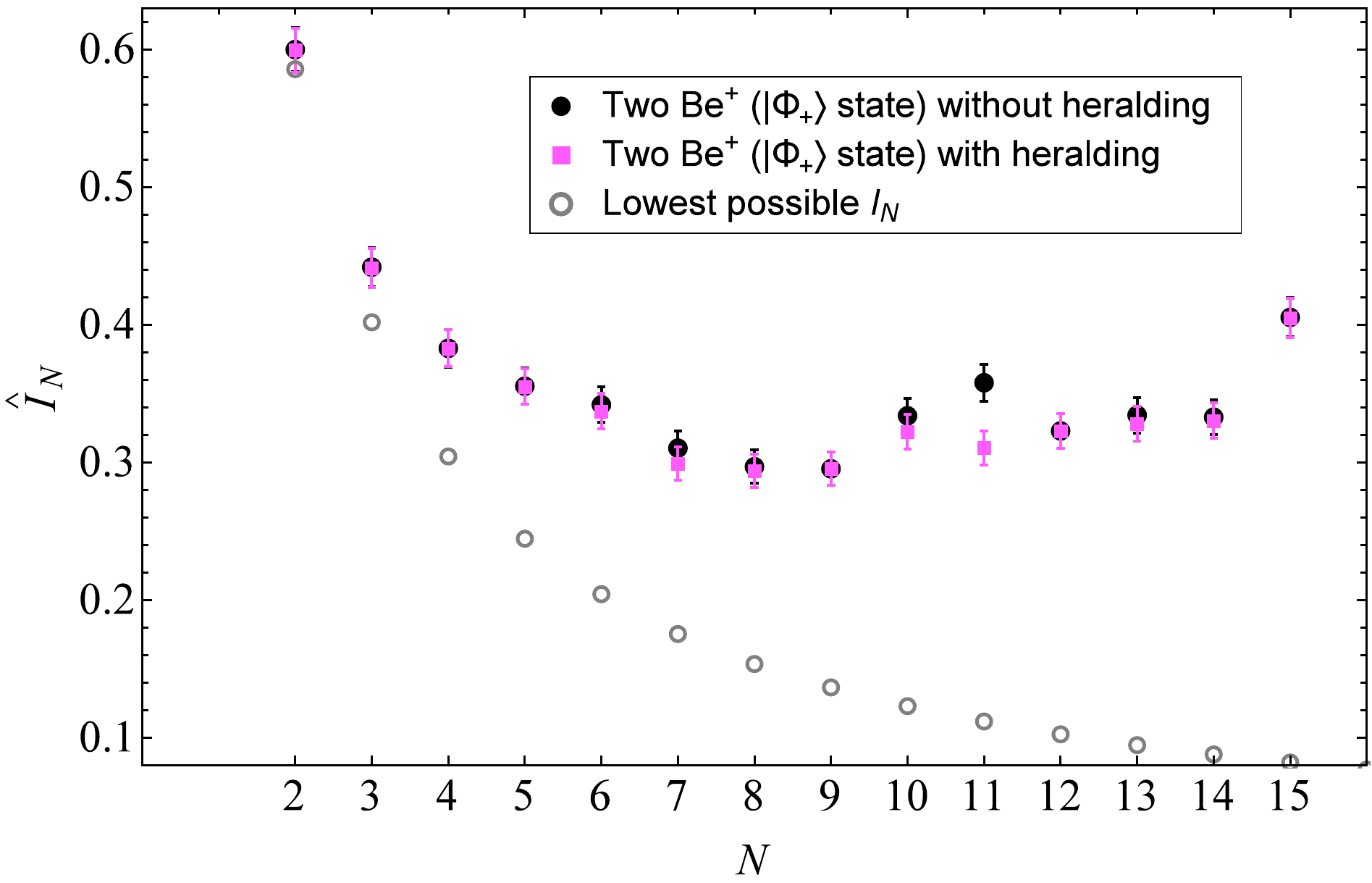}
\caption{Experimentally measured values of $\widehat{I}_N$ as a function of $N$ with and without the heralding collections of trials. Black dots are the data collected with two $^9$Be$^+$ ions prepared in the $\ket{\Phi_{+}}$ state (see Fig. \ref{ChainedBellTest}). }
\label{ChainedBellTestSimulated}
\end{figure}

Table \ref{Chained3} shows the frequency of outcomes for the $N = 3$ experiment with the $\ket{\Phi_-}$ state for each setting pair. Table \ref{Chained8} shows the averaged correlations for $N = 8$ with the $\ket{\Phi_+}$ state. Figure \ref{ChainedBellTestSimulated} shows the measured $\widehat{I}_N$ with and without using the heralding procedure to collect trials obtained with two $^9$Be$^+$ ions prepared in the $\ket{\Phi_{+}}$ state. 

\begin{figure}
\includegraphics[width=1\linewidth]{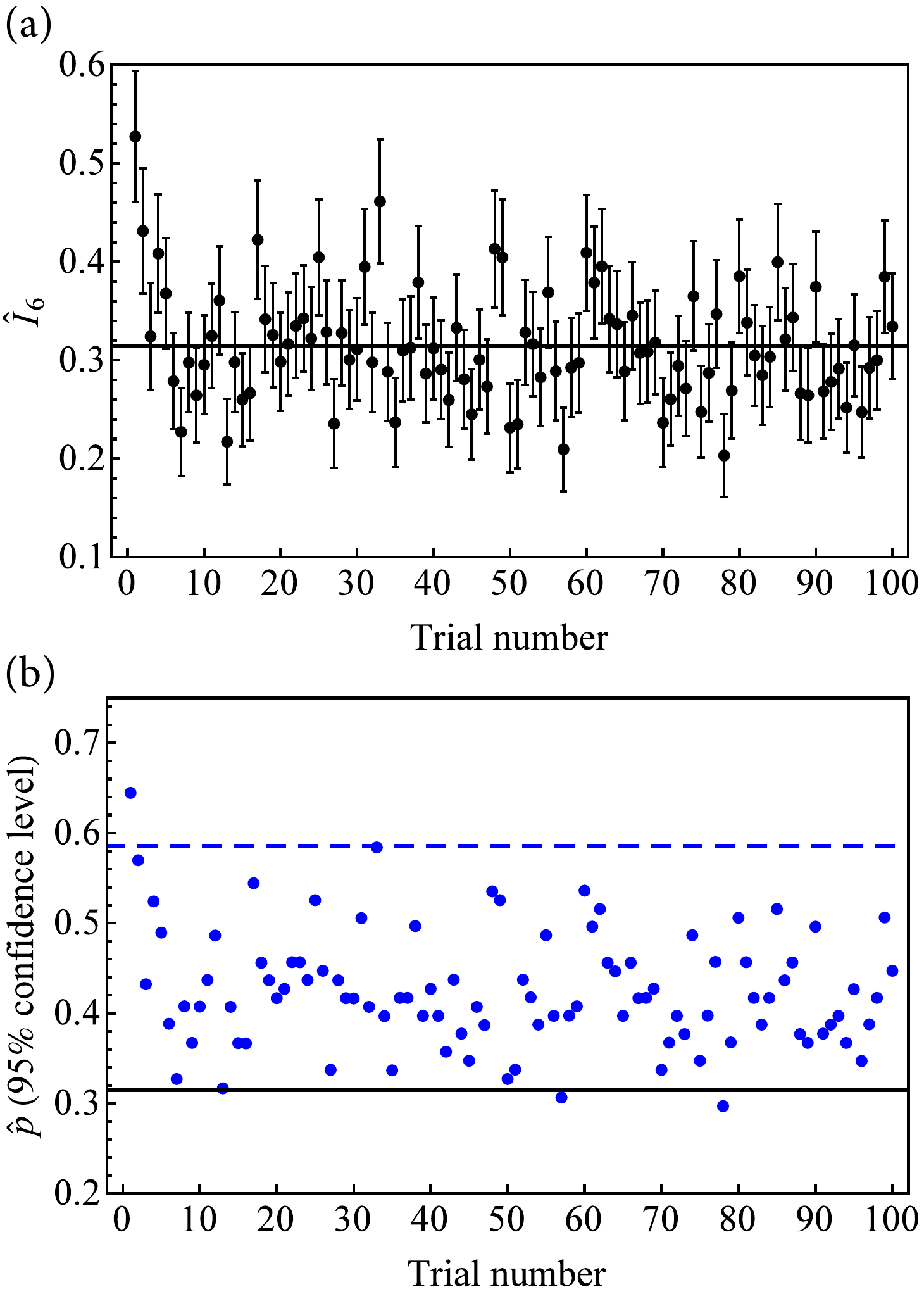}
\caption{(a) The computed CBI parameter, $\widehat{I}_6$ for all heralded data collected according to Eq. (\ref{FluorescenceCheck}) are plotted against the trial number within each block for the data run where settings are randomized. Error bars for $\widehat{I}_6$ are the standard error of the mean. The aggregate value is determined to be $\widehat{I}_6=0.315(5)$ (black solid line) when including all the data that pass the fluorescence check procedures. Trials having small trial number show less violation of the CBI; this is in part because fewer fluorescence detections were performed before those trials so that the check is not as rigorous as the checks for greater trial numbers. (b) The 95 \% confidence interval upper limit $\widehat{p}$. The blue dashed line indicates the lowest upper bound on $\plocal$ attainable using the CHSH inequality. }
\label{RandomIndividual}
\end{figure}

Table \ref{RandomizedResults} shows the frequencies of outcomes from the 1,361 successfully heralded trials obtained from the 50th trial in each block. Although the same settings are used for the previous 49 trials in each block, we assume that the ions and their environment retain no record of the settings at the beginning of the 50th trial. Ideally, for our statistical analysis, the number of successful trials used should be determined beforehand, because the reported confidence levels are valid only for a fixed number of trials. If the number of trials is not fixed in advanced, confidence level calculations would need to account for the possibility that the number of trials and the measurement outcomes could be dependent random variables. Here we use all available trials from a single day's experiment run. To improve the experiments, we would use the first trial for the main analysis with a fluorescence check performed before the trial for heralding. This would prevent the possibility that the settings applied during the first 49 trials in a block could influence the state of the ions during the 50th trial. In Table \ref{RandomizedResults} we also show the results from all (100) trials. Figure \ref{RandomIndividual} shows the CBI parameter, $\widehat{I}_6$ and the upper bound on the local part $\widehat{p}$ at the 95\% confidence level plotted as a function of trial number for the same (randomized settings) data run. 

\begin{table}
\begin{tabular}{|c|c|c|c|c|c|}
  \hline
        $a_k$ & $b_l$ & $BB$ & $BD$ & $DB$ & $DD$\\
        \hline
        $\pi/6$ & $0$ & $0.4799$ & $0.03296$ & $0.03403$ & $0.4531$ \\
        $\pi/6$ & $-\pi/3$ & $0.4705$ & $0.04700$ & $0.03858$ & $0.4439$ \\
        $\pi/2$ & $-\pi/3$ & $0.4666$ & $0.03458$ & $0.04104$ & $0.4577$ \\
        $\pi/2$ & $-2\pi/3$ & $0.4747$ & $0.04227$ & $0.03825$ & $0.4448$ \\
        $5\pi/6$ & $-2\pi/3$ & $0.4697$ & $0.03578$ & $0.03539$ & $0.4592$ \\
        $5\pi/6$ & $0$ & $0.05345$ & $0.4437$ & $0.4614$ & $0.04140$ \\
        \hline
\end{tabular}
\caption{Frequencies for observing outcomes $BB$, $BD$, $DB$, and $DD$ for each of the measurement settings used in an $N=3$ data run with the state $\ket{\Phi_-}=\frac{1}{\sqrt{2}}\left(\left|\uparrow\uparrow\right>-\left|\downarrow\downarrow\right>\right)$.  Frequencies are determined from 2,500 trials for each setting pair $a_kb_l$. Trials with different settings were implemented in blocks of 250, but not randomized. } 
\label{Chained3}
\end{table}

\begin{table}
\begin{tabular}{|c|c|c|}
  \hline
  $a_k$ & $b_l$ & $\overline{\mathcal{C}}(a_k,b_l)$\\
  \hline
  $\pi/16$ & $0$ & $0.0175(29)$\\
        $\pi/16$ & $-\pi/8$ & $0.0265(36)$\\
        $3\pi/16$ & $-\pi/8$ & $0.013(25)$\\
        $3\pi/16$ & $-2\pi/8$ & $0.022(33)$\\
        $5\pi/16$ & $-2\pi/8$ & $0.009(21)$\\
        $5\pi/16$ & $-3\pi/8$ & $0.018(30)$\\
        $7\pi/16$ & $-3\pi/8$ & $0.018(30)$\\
        $7\pi/16$ & $-4\pi/8$ & $0.0195(31)$\\
        $9\pi/16$ & $-4\pi/8$ & $0.0135(26)$\\
        $9\pi/16$ & $-5\pi/8$ & $0.022(33)$\\
        $11\pi/16$ & $-5\pi/8$ & $0.0155(28)$\\
        $11\pi/16$ & $-6\pi/8$ & $0.021(32)$\\
        $13\pi/16$ & $-6\pi/8$ & $0.0205(32)$\\
        $13\pi/16$ & $-7\pi/8$ & $0.0245(35)$\\
        $15\pi/16$ & $-7\pi/8$ & $0.019(31)$\\
        $15\pi/16$ & $0$ & $0.9825(29)$\\
  \hline
\end{tabular}
\caption{Settings and the averaged correlation outcomes for $\widehat{I}_8$ with the state $\ket{\Phi_+}=\frac{1}{\sqrt{2}}\left(\left|\uparrow\uparrow\right>+\left|\downarrow\downarrow\right>\right)$. For each setting pair, $M=2,000$ trials were carried out before changing to the next setting pair. The standard error $\sigma$ of $\overline{\mathcal{C}}$ for each settings pair is shown in parentheses, and is determined according to $\sigma=\sqrt{\mathcal{\overline{C}}(1-\mathcal{\overline{C}})/(M-1)}$. }
\label{Chained8}
\end{table}

\begin{table*}
\begin{tabular}{|c|c|c|c|c|c|c|c|c|c|c|c|}
  \hline
        & & \multicolumn{5}{c|}{50th trial} & \multicolumn{5}{|c|}{All trials}\\ \cline{3-12}
  $a_k$ & $b_l$ & No. of trials & $BB$ & $BD$ & $DB$ & $DD$ & No. of trials & $BB$ & $BD$ & $DB$ & $DD$ \\
        \hline
  $\pi/12$ & $0$ & $117$ & 1 & 52 & 62 & 1 & 11,650 & 157 & 5,538 & 5,745 & 210 \\
        $\pi/12$ & $-\pi/6$ & $114$ & 0 & 53 & 61 & 0 & 11,379 & 122 & 5,435 & 5,696 & 126 \\
        $3\pi/12$ & $-\pi/6$ & $106$ & 2 & 49 & 52 & 3 & 10,559 & 144 & 4,993 & 5,257 & 165 \\
        $3\pi/12$ & $-2\pi/6$ & $97$ & 1 & 47 & 48 & 1 & 9,690 & 90 & 4,566 & 4,919 & 115 \\
        $5\pi/12$ & $-2\pi/6$ & $107$ & 0 & 57 & 50 & 1 & 10,675 & 148 & 4,990 & 5,390 & 147 \\
        $5\pi/12$ & $-3\pi/6$ & $118$ & 1 & 55 & 62 & 0 & 11,859 & 115 & 5,621 & 5,952 & 171 \\
        $7\pi/12$ & $-3\pi/6$ & $136$ & 3 & 65 & 67 & 0 & 13,554 & 192 & 6,443 & 6,723 & 196 \\
        $7\pi/12$ & $-4\pi/6$ & $119$ & 0 & 57 & 60 & 2 & 11,884 & 110 & 5,641 & 5,972 & 161 \\
        $9\pi/12$ & $-4\pi/6$ & $120$ & 0 & 55 & 64 & 1 & 11,987 & 205 & 5,751 & 5,884 & 147 \\
        $9\pi/12$ & $-5\pi/6$ & $111$ & 0 & 56 & 55 & 0 & 11,014 & 91 & 5,430 & 5,340 & 153 \\
        $11\pi/12$ & $-5\pi/6$ & $113$ & 4 & 52 & 54 & 3 & 11,295 & 203 & 5,505 & 5,435 & 152 \\
        $11\pi/12$ & $0$ & $105$ & 42 & 0 & 3 & 59 & 10,461 & 5,218 & 86 & 174 & 4,983 \\
  \hline
\end{tabular}
\caption{Settings and the outcomes for the 50th trial and for all trials of the $N = 6$ randomized settings data run. }
\label{RandomizedResults}
\end{table*}

\subsection{Summary of high-efficiency measurement CHSH experiments}

Table \ref{tab:others} summarizes the $B_\text{CHSH}$ parameters determined in several previous detection-loophole closing experiments and ours, as well as the resulting self-testing singlet-fidelity lower-bound according to Ref.~\cite{Kaniewski2016} at the 50 \% and 95 \% confidence levels. The fidelity bounds $F_{\text{l}}$ are calculated with 
\begin{equation}
F_{\text{l}}=\frac{1}{2}\left(1+\frac{B_{\text{CHSH}}-\beta_S}{2\sqrt{2}-\beta_S}\right),
\end{equation}
where $\beta_S=(16+14\sqrt{2})/17$.  For the 50 \% confidence fidelity lower bound we use the point estimates of $B_{\text{CHSH}}$ given in Table~\ref{tab:others}, and for the 95 \% confidence lower bound we replace $B_\text{CHSH}$ in the equation above with a 95 \% confidence lower bound on $B_{\text{CHSH}}$ assuming that each estimate of $B_{\text{CHSH}}$ is normally distributed with standard deviation given by the uncertainties in the table. Table IV lists only experiments reporting violation of the CHSH inequality. Other experiments have closed the detection loophole, but they reported violation of other Bell inequalities, for which self-tested fidelity bounds are not available.

\begin{table}
\caption{Results from CHSH experiments without the fair-sampling assumption. The table shows each experiment's measured CHSH parameter $B_\text{CHSH}$ with one standard deviation uncertainty from the references and self-testing fidelity lower bounds at the 50 \% and 95 \% confidence levels determined as described in the text.}
\label{tab:others}
\begin{ruledtabular}
\begin{tabular}{ccddd}
& & & \multicolumn{2}{c}{Fidelity bounds} \\
 & System & \multicolumn{1}{c}{$B_{\text{CHSH}}$} & \multicolumn{1}{c}{50 \%} & \multicolumn{1}{c}{95 \%} \\
\hline
\cite{Rowe2001} & Two \tsup{9}Be\tsup{+} \footnote{Two \tsup{9}Be\tsup{+} ions confined in the same well of an ion trap, measured jointly.  This was the first experiment to close the detection loophole.} & 2.25(3) & 0.600 & 0.566 \\
\cite{Matsukevich2008} & Two \tsup{171}Yb\tsup{+} \footnote{Two \tsup{171}Yb\tsup{+} confined in traps separated by $~1$ m.  The ions were entangled by swapping entanglement with photons.} & 2.54(2) & 0.800 & 0.778 \\
\cite{Ansmann2009} & Phase qubits \footnote{Josephson phase qubits coupled by a coplanar waveguide.} & 2.0732(3) & 0.477 & 0.477\footnote{With more digits these two results are 0.47743 at 50 \% and  0.47709 at 95 \%.}\\
\cite{Pironio2010} & Two \tsup{171}Yb\tsup{+} \footnote{Similar to Ref. \cite{Matsukevich2008}.} & 2.414(58) & 0.713 & 0.647 \\
\cite{Hofmann2012} & Two \tsup{87}Rb\tsup{0} \footnote{Two neutral \tsup{87}Rb atoms confined in traps separated by $~20$ m. The ions were entangled by swapping entanglement with photons.} & 2.19(9) & 0.558 & 0.456 \\
\cite{Pfaff2013} & One NV \footnote{Two nuclear spins associated with a single nitrogen-vacancy center in diamond.} & 2.30(5) & 0.634 & 0.577 \\
\cite{Vlastakis2015} & Trans. \& cavity \footnote{A transmon qubit was entangled with a microwave cavity. The same cavity was used to measure the qubit.} & 2.30(4) & 0.634 & 0.589 \\
\cite{Ballance2015} & \tsup{40}Ca\tsup{+} \& \tsup{43}Ca\tsup{+} \footnote{One \tsup{40}Ca\tsup{+} and one \tsup{43}Ca\tsup{+} ion confined in the same well of an ion trap.} & 2.228(15) & 0.584 & 0.567 \\
\cite{Tan2015} & \tsup{9}Be\tsup{+} \& \tsup{25}Mg\tsup{+} \footnote{One \tsup{9}Be\tsup{+} and one \tsup{25}Mg\tsup{+} ion confined in the same well of an ion trap.} & 2.70(2) & 0.911 & 0.888 \\
\cite{Dehollain2016} & One \tsup{31}P in Si \footnote{Entanglement of the electron and nuclear spin of a single \tsup{31}P atom in Si.} & 2.70(9) & 0.911 & 0.809 \\
\cite{Hensen2016} & Two NV \footnote{Two nitrogen-vacancy centers in diamond, separated by 1.3 km. The NV centers were entangled by swapping entanglement with photons. This experiment closed all loopholes in a CHSH test of local realism.}& 2.38(14) & 0.690 & 0.530 \\
This & Two \tsup{9}Be\tsup{+} & 2.80(2) & 0.980 & 0.958 \\
\end{tabular}
\end{ruledtabular}
\end{table}

\subsection{Statistical procedure and proof of claims}

Here we show that $[0,\widehat{p}]$ is a $1-\alpha$ confidence interval for $\pml$, paying careful attention to the precise definition of ``confidence interval'' due to the non-standard nature of our statistical procedure. This framework is different than the null-hypothesis test usually employed in experiments aiming to falsify local realism.

Recall the statistic (random variable) $T_i$ that we defined in the main text, Eq.~(\ref{eq:ti}). The values of $T_{i}$ are tabulated in Table \ref{Cdef}.
\begin{table}[h]
\begin{tabular}{|c|c|c|c|c|c|}
  \hline
        $a$ & $b$ & $BB$ & $BD$ & $DB$ & $DD$\\
        \hline
        $a_1$ & $b_1$ & 0 & 1 & 1 & 0\\
        $a_1$ & $b_2$ & 0 & 1 & 1 & 0\\
        $a_2$ & $b_2$ & 0 & 1 & 1 & 0\\
        $a_2$ & $b_3$ & 0 & 1 & 1 & 0\\
        \vdots &\vdots &\vdots &\vdots &\vdots &\vdots\\
        $a_N$ & $b_N$ & 0 & 1 & 1 & 0\\
        $a_N$ & $b_1$ & 1 & 0 &  0  & 1 \\
        \hline
\end{tabular}
\caption{The statistic $T_i$ for $\ket{\Phi_{+}}$ takes the values in the table above depending on the settings chosen and the measurement outcomes.}
\label{Cdef}
\end{table}
We consider an experiment that randomly selects settings pairs only from those that appear in the CBI (for example, $a_1b_3$ is never measured). If the random choice between the $2N$ admissible settings configurations is equally probable and independent of the state being measured, the CBI is equivalent to $P(T_i=1) \leq (2N-1)/(2N)$ for a local distribution. A nonlocal distribution can violate this inequality in the absence of signaling, so for a trial $i$ governed by a distribution with a local part of probability
$p^i_{\mathrm{local}}$, we have 
\begin{eqnarray}
P(T_i=1) &=& P(T_i=1|\text{local dist.})P(\text{local dist.})\nonumber\\
&& + P(T_i=1|\text{nonlocal dist.})P(\text{nonlocal dist.})\nonumber\\
&\le&\frac{2N-1}{2N}p^i_{\mathrm{local}} + 1(1-p^i_{\mathrm{local}})\nonumber\\
&=& \frac{2N - p^i_{\mathrm{local}}}{2N}.
\end{eqnarray}
In the presence of possible memory effects, $p^i_{\mathrm{local}}$ can change from trial to trial and can even be correlated with the outcomes of earlier trials. Thus we consider the minimum possible $p^i_{\mathrm{local}}$ that can occur over the course of the experiment, $\pml$ (satisfying $0\le \pml\le 1$ by definition), and we study the constraints imposed by different possible values of $\pml$. Crucially, the following bound holds 
\begin{equation}
\label{pastconditional}
P(T_i=1|T_1,..,T_{i-1})\le \frac{2N-\pml}{2N},
\end{equation}
because even though information about past outcomes might theoretically be correlated with the value of $p^{i}_{\mathrm{local}}$, it cannot decrease $p^{i}_{\mathrm{local}}$ below the minimum possible value $\pml$ (which may be zero). Now consider the following result, which is proved as the second proposition in Appendix C of \cite{Bierhorst2015}

\medskip

\noindent {\it Proposition.} Let $(T_i)_{i=1}^{n}$ be a sequence of random variables taking values in the set $\{0,1\}$. Suppose that there exists a number $q\in (0,1)$ such that $\forall i$, $P(T_i=1|T_1,...,T_{i-1})\le q$. Then for any $y \in \{0,1,...,n\}$, the following holds
\begin{equation}
\label{binombound}
P(T \geq y) \le B_{\mathrm{tail}}(y, n, q),
\end{equation}
where $T:=\sum_{i=1}^nT_i$ and $B_{\mathrm{tail}}(y, n, q)$ is the probability that a binomial random variable of $n$ trials and success probability $q$ takes a value greater than or equal to $y$.

\medskip

With this result, the desired confidence interval can be obtained by inversion of hypothesis-test acceptance regions (\S 7.1.2 of~\cite{Shao2003}), which we now summarize.  The proposition implies that the hypothesis $\pml=x$ (or $\pml\geq x$) can be tested at significance level $\alpha$ with the statistic $T$: Find the threshold $y(\alpha,x)$ such that
$y(\alpha,x) = \min_{y}\{y: B_{\mathrm{tail}}(y,n,(2N-x)/(2N)) \leq \alpha\}$ and reject the hypothesis if the observed value $t$ of $T$ satisfies $t\geq y(\alpha,x)$. Note that $y(\alpha,x)$ is continuous and strictly decreasing with $x$. Consequently, the acceptance region for this hypothesis test is an interval $[0,\widehat p)$, where $\widehat p$ is the minimum value of $x$ such that $y(\alpha,x)\leq t$, so we can use $[0,\widehat p]$ as a level-$\alpha$ confidence set.  Thus $\widehat{p}$ is given by
\begin{equation}
\widehat{p} = \min_x \left\{x:B_{\mathrm{tail}}\left(t,n,\frac{2N-x}{2N}\right)\leq \alpha\right\},
\end{equation}
or equivalently,
\begin{equation}\label{phatdef}
\widehat{p} = \max_x\left\{x:B_{\mathrm{tail}}\left(t,n,\frac{2N-x}{2N}\right)\ge \alpha\right\}.
\end{equation}
Note that $\widehat{p}$ is a sample from the random variable $\widehat{P}$, which is a function of the random variable $T$.

For completeness and as an example, we end this section by directly establishing the validity of our confidence intervals. According to the definition in \S 2.4.3 of~\cite{Shao2003}, $[0,\widehat{p}]$ is an $\alpha$-{\it level confidence set} for $\pml$ if the following statement holds for any probability distribution $P$ satisfying Eq. (\ref{pastconditional}) and governing the experiment
\begin{equation}
\label{CIdef}
P\left(\pml \in [0, \widehat{P}]\right) \ge 1-\alpha.
\end{equation}

To demonstrate that our definition of $\widehat{P}$ obeys Eq. (\ref{CIdef}), first consider the integer $y_{\text{min}} = \min_{y\in\mathbb N}\{y: P(T\ge y+1)\le \alpha\}$. Since $y_{\text{min}}$ satisfies $P(T\ge y_{\text{min}}+1)\le \alpha$, and since $P(T\ge y_{\text{min}}+1)+P(T\le y_{\text{min}})=1$ (as $T$ only takes integer values), the following holds
\begin{equation}
\label{trivial}
1-\alpha \le P\left(T\le \min_{y\in\mathbb N}\{y: P(T\ge y+1)\le \alpha\}\right).
\end{equation}
Now by Eq. (\ref{binombound}), $B_{\mathrm{tail}}(y+1,n, (2N-\pml)/(2N))\le \alpha$ implies $P(T\ge y+1)\le \alpha$, so every element of the set $\{y\in\mathbb N:B_{\mathrm{tail}}(y+1,n, (2N-\pml)/2N)\le \alpha\}$ is also an element of the set $\{y\in\mathbb N: P(T\ge y+1)\le \alpha\}$, and consequently
\begin{multline}
\min_{y\in\mathbb N}\{y: P(T\ge y+1)\le \alpha\} \le \\ 
\min_{y\in\mathbb N}\{y:B_{\mathrm{tail}}(y+1,n, (2N-\pml)/2N)\le \alpha\}.
\end{multline}
For any numbers $y_1$ and $y_2$, $y_1\le y_2$ implies $P(T\le y_1)\le P(T \le y_2)$, so we can combine the above inequality with Eq. (\ref{trivial}) to infer
\begin{multline}
\label{Tless}
1-\alpha\le\\
P\left(T\le \min_{y\in\mathbb N}\{y:B_{\mathrm{tail}}(y+1,n, (2N-\pml)/2N)\le \alpha\}\right).
\end{multline}
Now if $T$ takes a value $t$ for which
\begin{equation}
t\le \min_{y\in\mathbb N}\{y:B_{\mathrm{tail}}(y+1,n, (2N-\pml)/2N)\le \alpha\},
\end{equation}
then $T$ has necessarily taken a value $t$ for which $B_{\mathrm{tail}}(t,n,(2N-\pml)/2N)$ is strictly greater than $\alpha$; otherwise, $t-1$ would have to be in the set $\{y:B_{\mathrm{tail}}(y+1,n, (2N-\pml)/2N)\le \alpha\}$ and so $t$ could not be less than or equal to the minimum of that set. Thus Eq. (\ref{Tless}) implies
\begin{equation}
\label{penultimate}
1-\alpha\le P\left(B_{\mathrm{tail}}(T,n,(2N-\pml)/2N)>\alpha\right).
\end{equation}
Finally, if we refer to the definition of $\widehat{p}$ in Eq. (\ref{phatdef}), we see that if $T$ takes a value $t$ for which $B_{\mathrm{tail}}(t,n,(2N-\pml)/2N)>\alpha$, then $\widehat{p}$, which is a function of $t$, satisfies $\widehat{p} \ge \plocal^{\mathrm{min}}$.  Hence Eq. (\ref{penultimate}) implies
\begin{equation}
1-\alpha\le P\left(\widehat{P} \ge \plocal^{\mathrm{min}}\right).
\end{equation}
This implies Eq. (\ref{CIdef}) and ensures that $[0,\widehat{p}]$ is an $\alpha$-level confidence set.

\end{document}